\documentclass[epjST]{svjour}
\usepackage{graphicx}
\usepackage{amssymb}
\newcommand{\beq}{\begin{equation}}
\newcommand{\eeq}{\end{equation}}
\newcommand{\beqs} {\begin{displaymath}}
\newcommand{\eeqs} {\end{displaymath}}
\newcommand{\beqa} {\begin{eqnarray}}
\newcommand{\eeqa} {\end{eqnarray}}
\newcommand{\beqas} {\begin{eqnarray*}}
\newcommand{\eeqas} {\end{eqnarray*}}

\begin{document}

\title{Competition of Brazil nut effect, buoyancy, and inelasticity induced
segregation in a granular mixture}

\author{Ricardo Brito\inst{1} \and Rodrigo Soto\inst{2}}

\institute{
Departamento de F\'{\i}sica Aplicada I and GISC, Facultad de Ciencias
F\'{\i}sicas, Universidad Complutense, 28040 Madrid, Spain. \and
Departamento de F\'{\i}sica, FCFM, Universidad de Chile, 
Santiago, Chile.}

\abstract{
It has been recently reported that a granular mixture in which grains differ
in their restitution coefficients presents segregation: the more inelastic
particles sink to the bottom. When other segregation mechanisms as buoyancy and the Brazil nut effect are present, 
the inelasticity induced segregation can compete with them. 
First, a detailed analysis, based on numerical simulations of two dimensional systems, of
the competition between buoyancy and the inelasticity induced segregation is
presented, finding that there is a transition line in the parameter space that determines which mechanism is dominant.
In the case of neutrally buoyant particles having different sizes the inelasticity
induced segregation can compete with the Brazil nut effect (BNE). Reverse Brazil
nut effect (RBNE) could be obtained at large inelasticities of the intruder. At
intermediate values, BNE and RBNE coexist and large inelastic particles are found both near the bottom and at the top of the system.
} 

\maketitle

\section{Introduction}

It is known from decades that when a mixture of two types of granular particles 
are externally excited by means, e.g. a vibration,  grains of different 
size, shape, mass or mechanical properties can mix or segregate.
For instance in a mixture of small grains and one large
intruder, when vertically vibrated, the intruder can go up \cite{Rosato} or down \cite{Breu} --the so called
direct and reverse Brazil nut effects, respectively. This effect hat has been studied in
many papers (see, e.g. Ref. \cite{Kudrolli} and references therein), where different 
mechanisms have been proposed to explain the segregation phenomenon. 
For example, percolation, arching, void filling, convection, interstitial 
fluid effects, condensation, and some more have been invoked to be the cause of segregation. 
All these effects are explained in Ref. \cite{Kudrolli}, although the problem is not completely 
understood yet. 
When both species have similar sizes (but possibly different) we can select few
cases where a variety of
segregation mechanisms and scenario appear \cite{RuizSuarez,Schroter,Khakhar}.
For instance, in Ref. \cite{Mullin}
particles of different masses, radii and restitution coefficients are placed in
a dish
which is horizontally vibrated, finding complete segregation. Segregation is
also found in the same geometry
when the grains have different friction coefficient with the base
\cite{Ciamarra}.
Under horizontal swirling, radial
segregation of particles of different sizes has been observed
\cite{Schnautz}. In avalanches, grains of different shape segregate in
stripes \cite{Makse}; in partially filled rotating drums, axial size
segregation develops \cite{Hill}. In two dimensional systems under gravity,
sinusoidally
vibrated, clustering has been observed \cite{King}.
This segregation effect can be modulated by using non-sinusoidal vibration
\cite{King2}.

In some of the cases mentioned above the grain species differ on the friction or
restitution coefficient. However few papers have studied segregation when
this is the only difference between grains. One of these cases is
Ref. \cite{Kondic}, where a mixture of spheres
that only differ in friction coefficients (static, dynamic and
rolling) is horizontally vibrated. They find complete mixing --that is, no
segregation-- for a flat plate
while segregation is only observed when the plate was slightly inclined.
Therefore, these results contradict the previously mentioned ones.

In a theoretical approach Ref. \cite{Serero} constructs the hydrodynamic
equations from the Boltzmann equation, finding inelasticity induced segregation.
The authors explain the phenomenon as a consequence of the
temperature
gradient in the system induced by inelastic collisions, and relate the
concentration gradient with the temperature gradient.
In the same spirit, Ref. \cite{Brey} studies the low density hydrodynamics of a
mixture in the so called
tracer limit, i.e. where the concentration of one of the components tends to
zero. Among other results, they find that the temperature ratio of both species
must be a constant. 
This constant value had been already measured by two experimental groups,
both in 3D \cite{Wildman} and 2D \cite{Menon} and by means of computer
simulations in 3D \cite{Talbot} and 2D \cite{Paolotti}. They found that the temperature ratio 
reaches a constant value in the regions of the system where the density is low. No tracer limit is needed. 
Generalization to moderate density has been done by Garz\'o \cite{Garzo} 
based on a kinetic
approach using the Enskog equation for dissipative hard spheres.

Recently, it has been shown numerically that in moderately dense vibrofluidized
granular matter, inelasticity induced segregation takes place
\cite{SegregaRestPRE} . When a mixture of grains of equal size and mass, but
differing in their restitution coefficient in put in a vibrating box, the more
inelastic grains sink, segregating partially with the other species. In this
paper we study if this inelasticity induced segregation can compete with two
other known mechanisms of grain segregation: buoyant forces when grain
densities differ and the Brazil nut effect when their sizes differ.

The structure of this paper is as follows. In Section \ref{sec.description} we
describe the system under consideration. Section \ref{sec.macroseg} reproduces
the results presented in Ref. \cite{SegregaRestPRE} regarding the macroscopic
segregation of two species differing only in their restitution coefficient.
Section \ref{sec.buoyant} analyzes the competition with buoyant
forces, when considering grains of equal size, but different masses and
restitution coefficients. Section \ref{sec.nonbuoyant} studies the competition
with the Brazil nut effect when grains of the two species have the same
density, but differ in size and restitution coefficients. 
We conclude with Section \ref{sec.conclusions} summarizing the results of the
paper.

\section{Description of the system} \label{sec.description}
We study the effect of the difference on restitution coefficients in
 the segregation phenomenon, by  means of Molecular Dynamics simulations of a
bidimensional  granular mixture of two types of particles,
$A$ and $B$. Grains are modeled as smooth Inelastic Hard Disks, but differing on
the
normal restitution
coefficient that characterizes their inelastic collisions. The two species can
also differ in their masses $m_A$ and $m_B$, and in their diameters $\sigma_A$
and $\sigma_B$.
The restitution coefficient for $A$-$A$ collisions is $\alpha_A$, for $B$-$B$
collisions is $\alpha_B$. For the interparticle collisions $A$-$B$  we have
taken
$\alpha_{AB}=(\alpha_A+\alpha_B)/2$. Usually we will consider
that B are the most inelastic particles ($\alpha_B<\alpha_A$).

We have taken a fixed total number of particles $N_T=N_A+N_B$, changing the
concentration of the $B$ particles.
For the simulations reported in this paper, we have fixed $N_T=680$ disks
and varied $N_B$ from 10 (that can be considered as a tracer limit)
until 160. The disks are placed under the action of a
gravitational acceleration $g$ pointing downward in a rectangular
box of width $L_x=50\sigma_A$, infinite height, and with the bottom wall
oscillating periodically at high frequency $\omega$ and small amplitude
$A$, with a bi-parabolic waveform \cite{Soto}. Periodic boundary conditions 
are used in the horizontal direction. They 
avoid the appearance of convective rolls by the influence of the walls.
Under these conditions, the system reaches a stationary
state with gradients in the vertical direction \cite{Grossman}.

Units are chosen such that $\sigma_A\equiv\sigma=1$, $m_A=1$ and $k_B=1$. 
The time scale is fixed  by the characteristic energy of wall oscillation, 
that is  $m_A(A\omega)^2=1$. Simulations are performed with a fixed value of
the gravity acceleration $g$ in order to reduce the parameter space and provide
a more detailed analysis  
on the effect of inelasticity. The value of  $g=0.15$ (in our units), 
 was chosen because  
the associated value of the ratio of the acceleration induced by the vibration
versus the gravity, $\Gamma=(A w^2)/g$ can be easily reached  
experimentally, as it takes the value of 
$\Gamma\simeq 6.67$. Finally, the amplitude on the vibration is set 
to $A=0.01\sigma$.

\section{Inelasticity induced segregation} \label{sec.macroseg}

In this section we reproduce the main results concerning the macroscopic
segregation of the two species when they only differ in restitution
coefficient, that were already reported in Ref. \cite{SegregaRestPRE}.
First, to illustrate the main observed features,  we report results of a
simulation having a small fraction of inelastic particles $N_B=10$, and
$\alpha_B=0.7$ and the rest nearly elastic: $N_A=670$, $\alpha_A=0.98$.

\begin{figure}[htb]
\centering\includegraphics[angle=0,clip=true,width=0.85\columnwidth,
keepaspectratio]{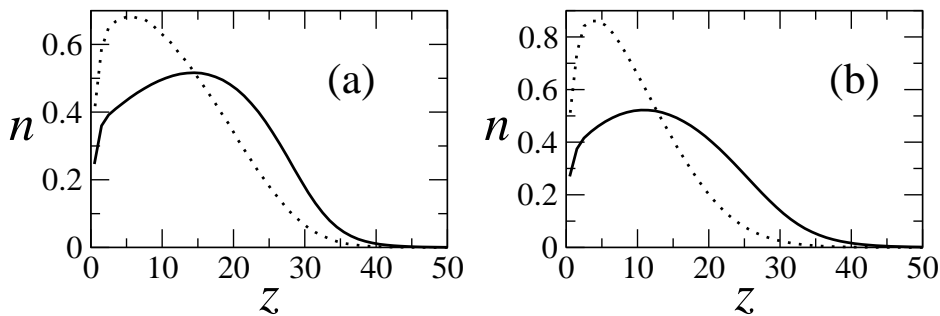}
\caption{Density profiles for the two species, density of $A$, $n_A$, (solid
line) and the rescaled density of $B$, $n_B N_A/N_B$, (dotted line). (a):
both species are inelastic with $N_A=670$, $N_B=10$, $\alpha_A=0.98$,
$\alpha_B=0.7$. (b): $A$ is elastic while $B$ is inelastic and $N_A=640$,
$N_B=40$, and $\alpha_B=0.5$}
\label{fig.densprofiles}
\end{figure}

The density profiles of the two species are shown in
Fig. \ref{fig.densprofiles}a, where we plot
the number density of particles of type $A$ and $B$: $n_A(z)$ and $n_B(z)$. The
normalization
of these quantities is such that $\int_0^\infty dz \int_0^{Lx}dx  n_A(z) = N_A$
(resp. $B$).
For plotting purposes only, $n_B$ is rescaled by a factor $N_A/N_B$, so in the
case of no segregation both profiles would be identical.
Both densities have the characteristic shape of
vibrofluidized systems subject to gravity: there is a initial density
increase due to the abrupt temperature drop caused by dissipation, and at
higher positions, density decreases again due to gravity
\cite{Grossman}. Density exhibits a maximum at $z\simeq 15\sigma$
where the density $n\simeq 0.5$, so the system cannot be considered as dilute.
The density profile of the more inelastic particles, $B$, is plotted as a dotted
line in Fig 1a.
Its maximum is located at smaller $z$,  indicating that they are closer to the
bottom of the container as compared to the more elastic ones. The segregation is
not complete, though.

\begin{figure}[htb]
\centering\includegraphics[angle=0,clip=true,width=0.85\columnwidth,
keepaspectratio]{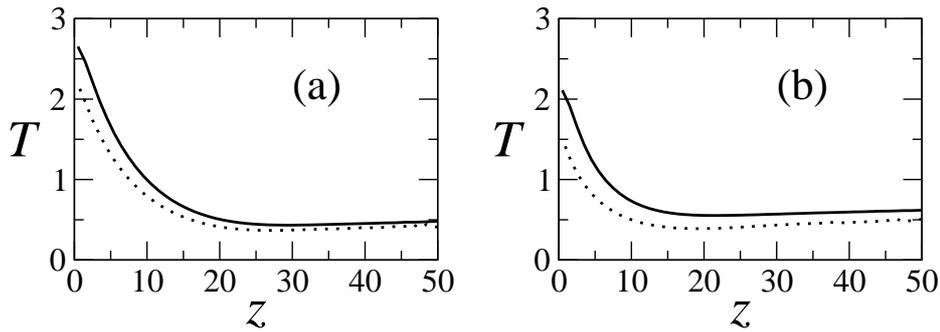}
\caption{Temperature profiles for two species, $T_A$ (solid
line) and $T_B$ (dotted line). (a) and (b) graphics have the same
parameters as in Fig. \ref{fig.densprofiles}}
\label{fig.tempprofiles}
\end{figure}

The temperature profiles are also highly inhomogeneous, as shown in Fig.
\ref{fig.tempprofiles}a. For both
species, the temperature presents a initial abrupt drop, but later (after
$z\simeq 20\sigma$) both profiles present a linear increase with height.
This phenomenon was already observed in one-component systems, and it is
associated to the energy transport term, $-\mu\nabla n$,  related to density
gradients, that appear in granular fluids \cite{RamirezSoto}. Let us note that
the maximum density does not coincide with the temperature minimum: there is a
shift between these two quantities which is qualitatively described by a
hydrostatic balance in presence of gravity \cite{RamirezSoto,Serero}.

The described segregation of species $A$ and $B$ is produced by their
different restitution coefficients as all the other properties are the same.
To study in more detail the effect of the difference of inelasticities, 
we proceed to
study the limiting case in which the $A$ particles are elastic
($\alpha_A=1$) and only the $B$ particles are
inelastic (and consequently, collisions $A$-$B$ are also inelastic).
In this way we also limit the parameter space, allowing to
a more detailed quantitative study.

Figures  \ref{fig.densprofiles}b and
\ref{fig.tempprofiles}b show the density and temperature profiles for such case,
where particles of type $A$ are elastic ($\alpha_A=1$) particles $B$ are
inelastic
($\alpha_B=0.5$) considering $N_B=40$.  It is observed
that the main properties of the profiles are preserved, even the positive
slope of $T_A$ despite the A-A collisions are elastic. Partial segregation is
again observed,
where inelastic particles, $B$, sink to the bottom of the container while
elastic
ones, $A$, are majority at upper layers of the fluid.

These results are not surprising in view of the predictions
of Ref.~\cite{Serero}, where it is argued that the segregation is
produced when the particles with different restitution coefficients are immersed
in a temperature gradient. The gradient is
induced by the inelastic collisions, so such gradient can be created vibrating a
mixture of elastic and inelastic particles. The latter ones dissipate some of
the energy
injected by vibration creating a stationary state. The hydrodynamic description
of the mixture also contains the dissipative flux $ -\mu \nabla n $, and
therefore it is expected that 
the hydrodynamic profiles of density and temperature will be equivalent to a
full inelastic system.

\begin{figure}[htb]
\centering\includegraphics[clip=true,width=0.55\columnwidth,
keepaspectratio]{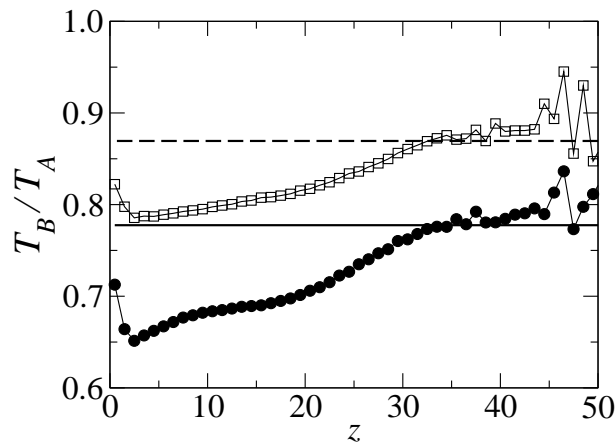}
\caption{Temperature ratios $T_B(z)/T_A(z)$ for the simulations
described in Fig. \ref{fig.tempprofiles}. Simulation results (solid circles) and
theoretical prediction in \cite{Brey} (solid line) for $N_A=670$, $N_B=10$,
$\alpha_A=0.98$, $\alpha_B=0.7$. Simulation results (open squares)
and
theoretical prediction in \cite{Brey} (dashed line) for A elastic,
$N_A=640$, $N_B=40$, and $\alpha_B=0.5$}
\label{fig.Tempratio}
\end{figure}

Figure \ref{fig.Tempratio} shows the temperature ratios $T_B(z)/T_A(z)$ for the
simulations
described in Fig. \ref{fig.tempprofiles}. At low
densities it was found
experimentally \cite{Wildman,Menon} and by employing kinetic theory \cite{Brey}
that such temperature ratio must be constant. However, we find a non constant
ratio
in the $z$ direction that only agrees with the result of \cite{Brey} at high
$z$, say for $z$ larger than 30. Inspection of Fig. \ref{fig.densprofiles}
shows that the density at $z>30$ is smaller than 0.2, so we can consider that 
the system is in a low density regime. Therefore the approximations of \cite{Brey}
are valid. 
In the case where A particles are elastic, equivalent predictions for the temperature ratio
were given in \cite{Martin,Trizac}.

In order to quantify the segregation, a series of simulations are performed with
$N_B$ ranging from 10 to 160, and $\alpha_B$ between 0.2 and 0.9. Larger values
of $N_B$ or
smaller restitution coefficients lead to clustering as described in
\cite{Meerson}.
For each simulation we compute the segregation parameter, defined as:
\beq 
\delta =1- 
 \int dz\, n_A(z)n_B(z)\bigg/ \sqrt{\textstyle\int dz\, n_A^2(z)\int dz\,
n_B^2(z)} \label{delta}
\eeq
where the $n_A(z)$ and $n_B(z)$ are the local density, as plotted in Fig.
\ref{fig.densprofiles}.
The segregation parameter is bounded between 0 and 1.  The value $\delta=1$ 
corresponds to complete segregation, as $\delta$ is 1 if $n_A(z)$ and $n_B(z)$
do not overlap. On the contrary, $\delta=0$ means complete mixing, as this value
can only be obtained if $n_B(z)$ is
proportional to $n_A(z)$. 

\begin{figure}[htb]
\centering\includegraphics[angle=0,clip=true,width=0.55\columnwidth,
keepaspectratio]{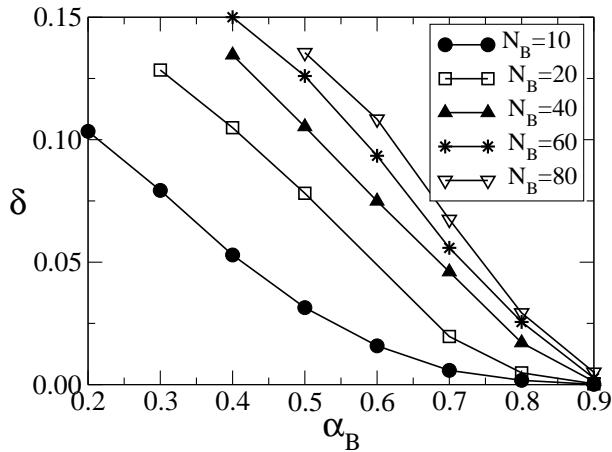}
\caption{Segregation parameter $\delta$ as a function of
$\alpha_B$ for $\alpha_{A}=1.0$ and $N_{T}=640$. Different curves correspond to various
concentrations of $B$ particles. In the elastic limit $\alpha\to 1$ all curves 
coincide at $\delta=0$.}
\label{fig.delta}
\end{figure}

The results for $\delta$ are collected in Fig. \ref{fig.delta} where the
quantity
$\delta$ is plotted versus the coefficient $\alpha_{B}$ for different values
of $N_B$. 
The fact that $\delta$ is always non vanishing
confirms that the segregation exists whenever the restitution coefficients are
different. Only in the case when $\alpha_B \to 1$ the quantity $\delta$
approaches 0, limit in which there is no segregation. Note that, for each
$\alpha_B$, $\delta$ increases with $N_B$. The results confirm that
segregation is not complete as $\delta$ never gets close to 1.
In addition, for each simulation, the center of mass of the $A$ and $B$
species are computed, $Z_{A/B}$, finding that $Z_A>Z_B$ in all cases.

The reported inelasticity induced segregation have a microscopic origin as
stated in Ref. \cite{SegregaRestPRE}: the most inelastic particles induce
locally a region of high density and low temperature, resembling a cold droplet
that falls in a gravitational field. The droplet is created by
the dissipation in a way that resembles the clustering instability
of the granular gases. Quantitative proof of this argument can be found in Ref. \cite{SegregaRestPRE},
by studying the density and temperature pair distribution functions
of the $A$ and $B$ particles.

\section{Competition with buoyancy forces} \label{sec.buoyant}

Up to now, we have considered a mixture where grains of both species have the same
masses and diameters. The main conclusion so far is that the inelasticity can create 
an effective force, pointing downwards, that sinks the more inelastic particles to
the bottom of the container. Besides, when densities differ, either because
the mass and/or the diameters of the particles are different, buoyancy forces appear 
making lighter grains move to the top. It can be asked whether the inelasticity induced
segregation could compensate the buoyancy force experienced by
lighter $B$ particles. To confirm or deny this idea, a series of
simulations with equal sized particles ($\sigma_B/\sigma_A=1$), keeping fixed
$\alpha_A$, $N_A$, and $N_B$, but varying $m_B/m_A$ and $\alpha_B$ 
is performed. In order to determine which force dominates, we compute the center of
masses of the $A$ and $B$ particles, denoting them by
$Z_A, Z_B$. Their precise definition is: 
\begin{equation}\label{Z}
Z_A=\frac{\int_0^\infty\, dz\, z n_A(z)}{\int_0^\infty\, dz\,  n_A(z)}
= \frac{\int_0^\infty\, dz\, z n_A(z)}{N_A/L_x},\ \ \mbox{(resp. $B$)}.
\end{equation}
If $Z_A > Z_B$ particles of type $A$ will be, on average, closer to
the top of the 
container than particles of type $B$ and viceversa. The ratio of $Z_B/Z_A$ being larger or smaller
than one will be our criterion for deciding which force actually dominates.  

\begin{figure}[htb]
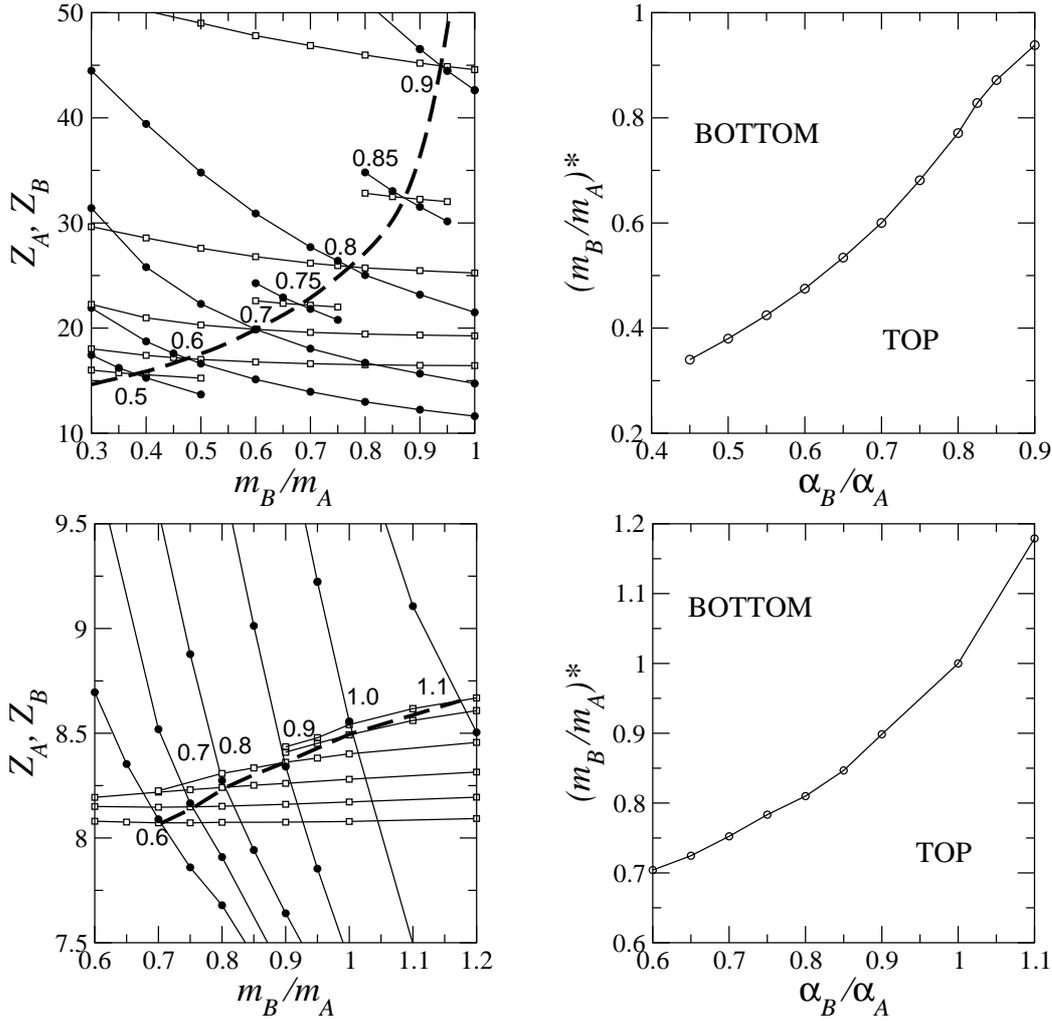

\centering\includegraphics[angle=0,clip=true,width=.95\columnwidth,
keepaspectratio]{Figura5.eps}
\centering\includegraphics[angle=0,clip=true,width=0.95\columnwidth,
keepaspectratio]{Figura6.eps}
\caption{Left: Position of the center of mass of $A$ (open squares) and $B$
particles (solid circles) as a function of the relative mass of $B$ particles
$m_B/m_A$ for different values of the ratio of the restitution coefficients, $\alpha_B/\alpha_{A}$ (indicated in the plot). 
The locus of the mass ratio at which the two centers of mass
coincide is indicated by a thick dashed line.
Right: Critical mass ratio $(m_B/m_A)^*$ at which the two centers of mass
coincide as a function of the ratio of the restitution coefficients
$\alpha_B/\alpha_{A}$. The regions
in the parameter space where $B$ particles move to the top or to the bottom
are indicated by the respective labels in the plot.
Simulation parameters are fixed to $N_{A}=640$, $N_{B}=40$,
$\sigma_B/\sigma_A=1$, and $\alpha_{A}=1$ (top) and
$\alpha_{A}=0.9$ (bottom).}
\label{fig.massratio1}
\end{figure}

The results for $\alpha_A=1$, $N_A=640$, and  $N_B=40$ are presented in Fig.
\ref{fig.massratio1}. The mass ratio $m_B/m_A$ 
is varied from 0.3 until 1, and the inelasticity of the $B$ particle from 0.4
until 0.9. For each value
of $\alpha_B$, there is a particular value of the mass
ratio, $(m_B/m_A)^*$ at which the positions of the two center of masses
coincide. The inelastic particles have lower center of mass if $m_B/m_A>
(m_B/m_A)^*$, and therefore sinking due to
inelasticity wins to the buoyancy force in this range. 
On the contrary, buoyancy force dominates if  $m_B/m_A < (m_B/m_A)^*$ and
lighter and inelastic particles go up. 

Figure \ref{fig.massratio1} also presents the value of the critical
ratio $(m_B/m_A)^*$ as a function of $\alpha_B$: more
inelastic particles have a higher tendency to sink and therefore a stronger
mass ratio is needed to compensate the inelasticity induced segregation. Note
that when $\alpha_B\to 1$, the positions of both center of masses diverge
because a vibrated elastic medium does not have a stationary state. There is an indication 
of such divergence  in the steep slope of the dashed line in the upper left panel of 
Fig. \ref{fig.massratio1}.
The segregation parameter $\delta$ does not vanish for any value of $m_B/m_A$,
indicating that there is no complete mixing even at the value of $m_B/m_A =
(m_B/m_A)^*$, where both center
of masses coincide. The value of $\delta$ (not shown here), however, is minimum at this precise
mass ratio. 
Also, $\delta$ does not reach 1 either, so no complete segregation is observed. 

A second series of simulations were done with $\alpha_A=0.9$ and 
are presented in Fig.  \ref{fig.massratio1}.  The qualitative features are similar to the 
the previous case ($\alpha_A=1.0$), except for some remarks. 
The first one in the center of mass of species $A$ is
about constant, independent of the mass ratio or the dissipation parameter of particles $B$. 
This is so because the particles $A$ are inelastic and dissipate a great deal of energy, forming a
very dense system. This density is  near the close packing, as it will be illustrated in the forthcoming figures. 
The second remark is that, as opposite to the $\alpha_A=1$ case, 
the center of masses do not diverge 
when $\alpha_B\to1$ due to the finite inelasticity of the $A$ particles. 
Finally, as $\alpha_A=0.9$, the ratio $\alpha_B/\alpha_A$ can exceed one. In particular, we can
make the $B$  particles elastic by choosing $\alpha_B=1/0.9=1.1111...$. We have done so and included the results 
in the bottom part of Fig. \ref{fig.massratio1}. As expected, particles $B$ must be heavier than 
$A$ in order to compensate the effective upwards force produced because they are more elastic. 
In this particular case the value of $(m_B/m_A)^*\simeq 1.17$.

\section{Competition with the Brazil nut effect}
\label{sec.nonbuoyant}
When grains are mixed such that they differ in size, a remarkable
phenomenon takes place: the larger grains rise to the top even if they are
denser. This phenomenon, called the Brazil nut effect (BNE), has been widely
studied in order to identify the different mechanisms that produce it and also
to characterize the range of parameters where it takes place. As mentioned in
the introduction, it has been found that besides the normal BNE, it is also
possible that, under certain conditions, the large particles sink in the
so-called reverse Brazil nut effect (RBNE).
In its simpler form, the BNE is studied in the neutrally buoyant case, where the
mass ratio between the species equals their volume ratio.  

\begin{figure}[b]
\centering \includegraphics[angle=270,clip=true,width=0.45\columnwidth,
keepaspectratio]{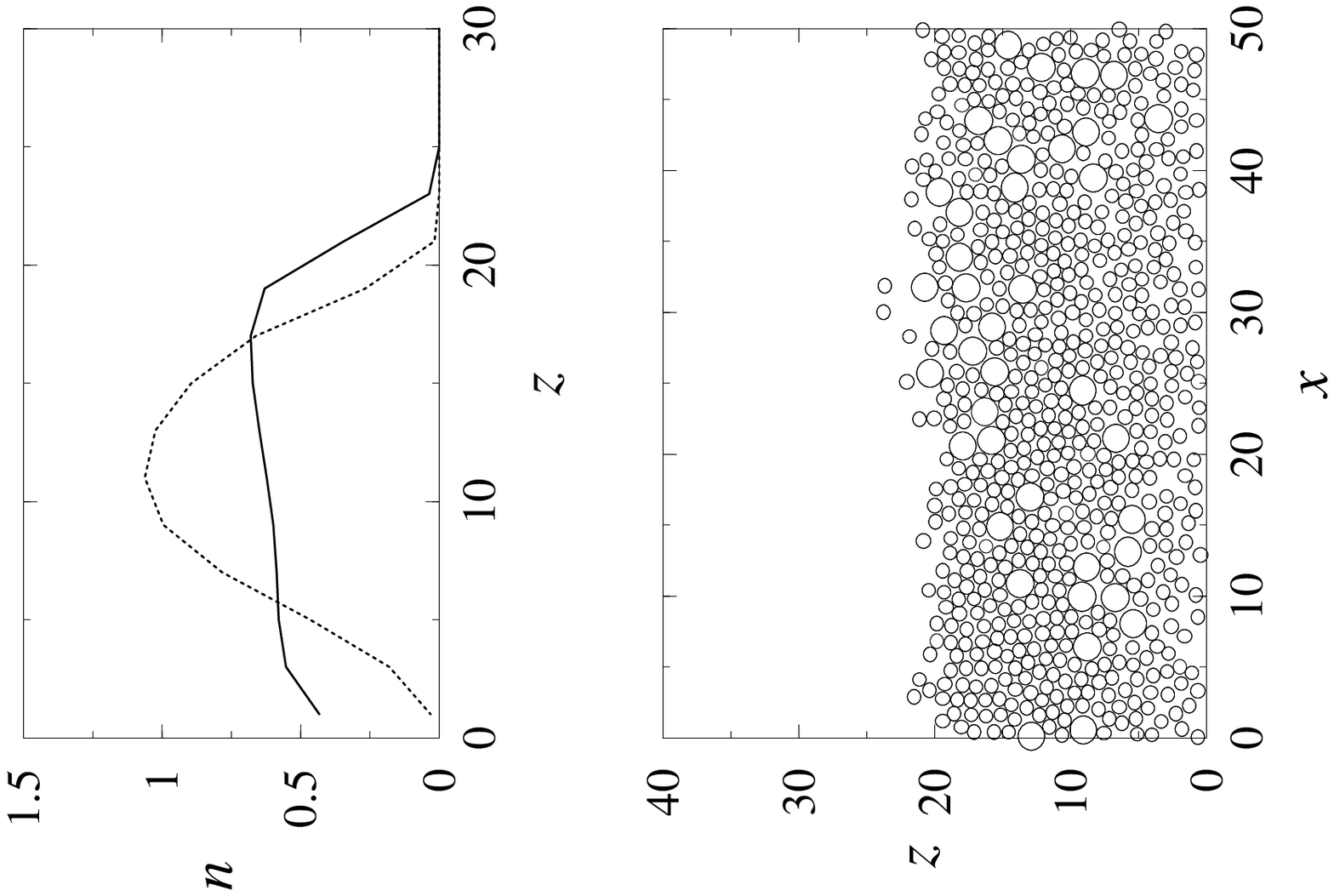}
\centering \includegraphics[angle=270,clip=true,width=0.44\columnwidth,
keepaspectratio]{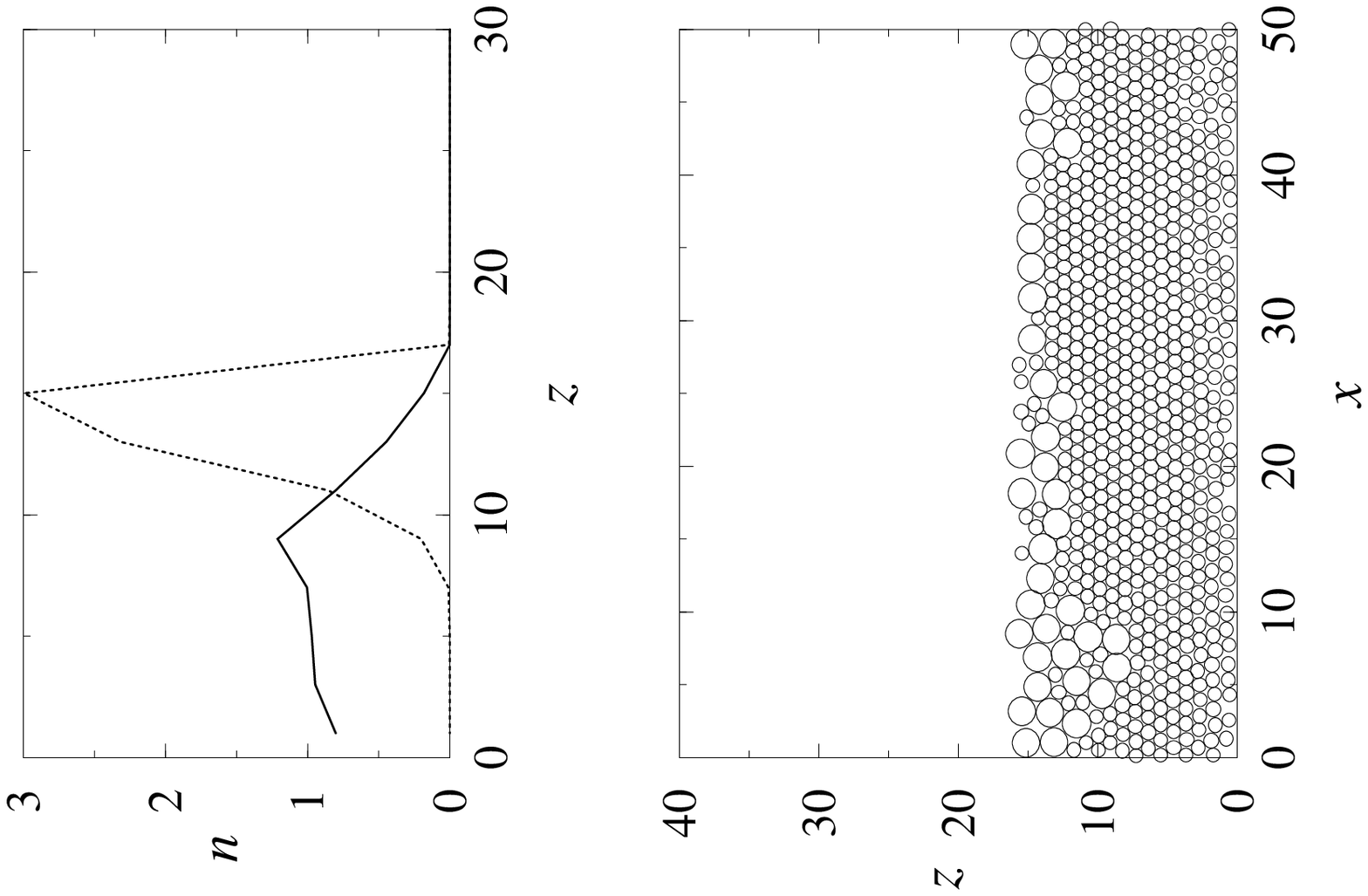}
\caption{Left: Top: Density profiles of $A$ particles (solid line) and $B$
particles
(dashed line) in a neutrally buoyant mixture with  $\sigma_B=2.0$, $N_A=640$,
$N_B=40$, and $\alpha_A=\alpha_B=0.95$ (left) and $\alpha_A=\alpha_B=0.85$
(right). Bottom: Configuration of the grains in the two systems.}
\label{fig.alphasigualesquasielastico}
\end{figure}

The results of the previous sections suggest that if the larger particle is also
more inelastic the inelasticity induced
segregation can compete with the BNE. We have performed a
series of simulations of a neutrally buoyant mixture (i.e.,
$m_B/m_A=(\sigma_B/\sigma_A)^2$) where we have varied the
size of $B$ particles $\sigma_B$
and the restitution coefficients $\alpha_A$ and $\alpha_B$, while the number of
particles are fixed to $N_A=640$ and $N_B=40$.
This parameter
space is huge as we have restricted to B particles that are not too large;
specifically we have considered $1\leq\sigma_B\leq 2.5$.
 
First, we consider the usual BNE case with $\alpha_A=\alpha_B$ and
$\sigma_B=1.5$, $2.0$ (that is, the two species only differ in their sizes). At
low dissipation ($\alpha_A=\alpha_B=0.95$) there is no BNE but rather
the two species mix with $B$ particles having a narrower distribution near the
center (see Fig. \ref{fig.alphasigualesquasielastico}). When the dissipation is
higher ($\alpha_A=\alpha_B=0.9,0.85$) the larger particles sit on the top as in
the usual BNE (see Fig. \ref{fig.alphasigualesquasielastico}). The segregation
is partial when $\sigma_B=1.5$ but it 
is complete when $\sigma_B=2.0$, differently as in the other cases,
where segregation was always partial. In summary, we have
checked that the BNE can be obtained with moderate size difference as long as
the inelasticity is strong enough.

\begin{figure}[htb]
\center\includegraphics[angle=0,clip=true,width=0.9\columnwidth,keepaspectratio]
{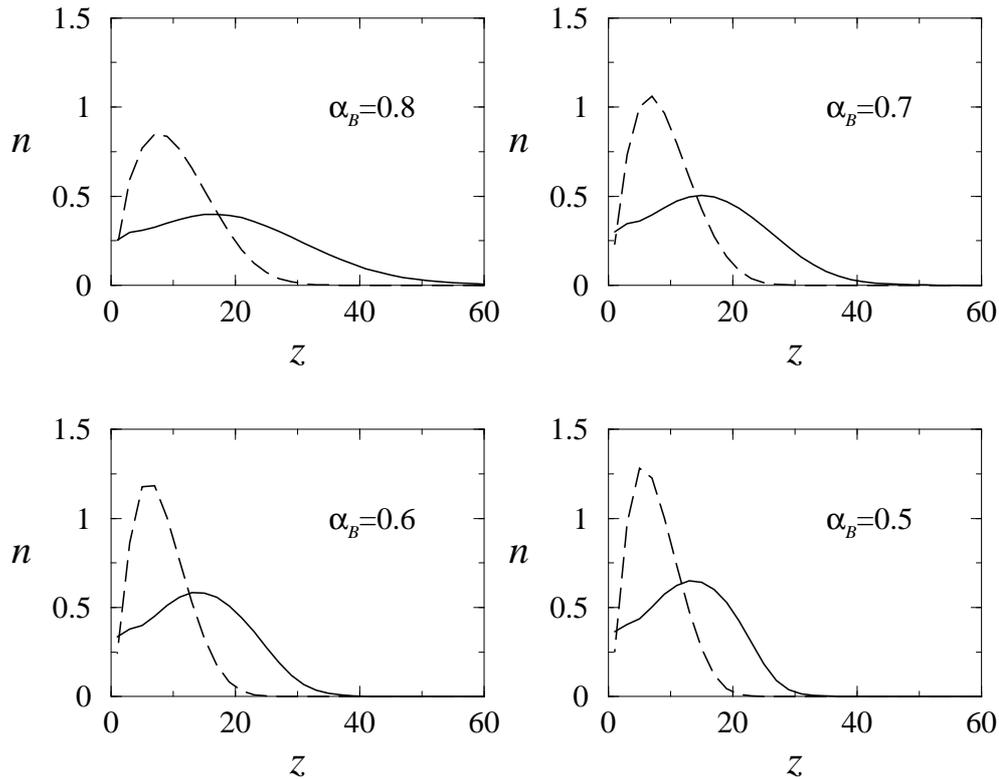}
\caption{Density profiles of $A$ particles (solid line) and $B$ particles
(dashed line) in a neutrally buoyant mixture with $\alpha_A=1$, $\sigma_B=2.0$,
$N_A=640$, $N_B=40$, 
and different values of $\alpha_B$ (indicated in the figures).}
\label{fig.alphasdiferenteselastico}
\end{figure}

When the two inelasticities differ (with $B$ particles being larger in size
and more inelastic), it is expected that the BNE could be compensated by the
inelasticity
induced segregation. A series of simulations with $\alpha_A=1$,
$\sigma_B=2.0$, and different values of $\alpha_B$ show that the inelasticity
induced segregation is the dominant mechanism and $B$ particles sink (that is,
RBNE is found) for large inelasticities and when the elastic limit is approached
both species mix (see Fig. \ref{fig.alphasdiferenteselastico}). However, the
elastic limit could not be achieved because, when both species are elastic, the
vibrated granular gas has no stationary state.

\begin{figure}[htb]
\center \includegraphics[angle=0,clip=true,width=0.90\columnwidth,keepaspectratio]
{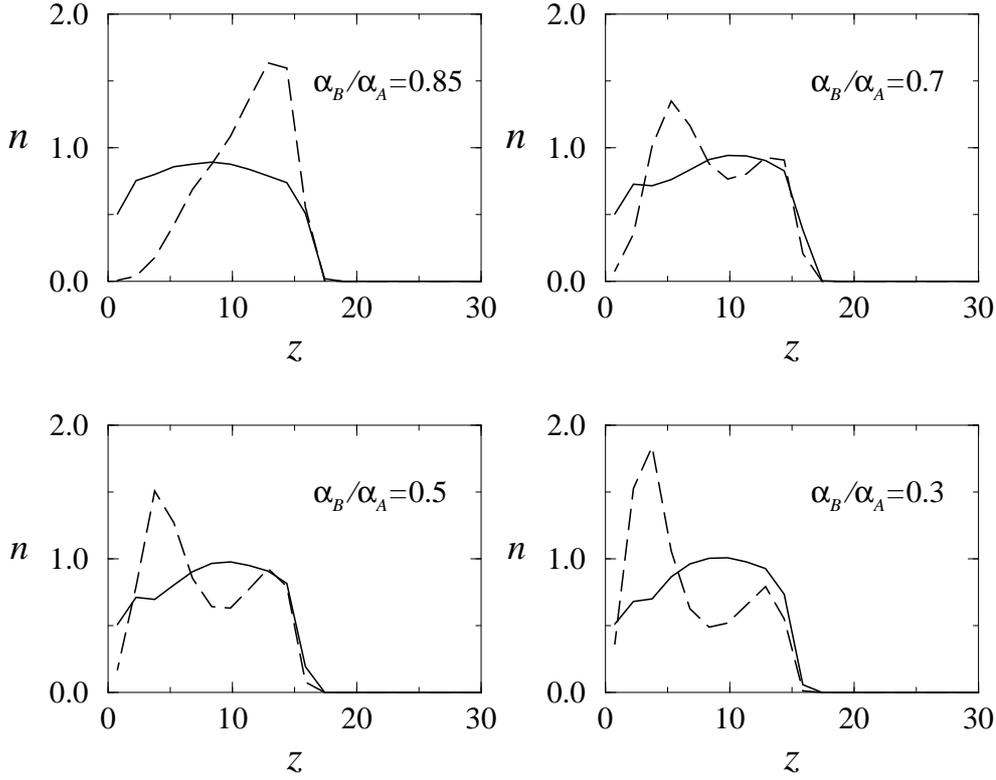}
\caption{Density profiles of $A$ particles (solid line) and $B$ particles
(dashed line) in a neutrally buoyant mixture with $\alpha_A=0.9$,
$\sigma_B=1.5$, $N_A=640$, $N_B=40$, and different values of $\alpha_B$
(indicated in the figures).}
\label{fig.RBNE-BNE}
\end{figure}

A different situation is found when $\alpha_A=0.9$ and $\sigma_B=1.5$, where a
transition from
RBNE to BNE is obtained with coexistence of both states. The density profiles
for different values of $\alpha_B$ are presented in Fig. \ref{fig.RBNE-BNE}.
When both inelasticities are equal or the $B$ particles are
slightly more inelastic than $A$ particles, a clear BNE is found. But when the
$B$
inelasticity is increased, an important fraction of $B$ particles sink showing a
coexistence of BNE and RBNE. In fact, a ``sandwich'' configuration is obtained
with $B$ particles being in the bottom and top layers while $A$ particles are in
the middle region. The segregation, as is common in all the examples presented
here, is not complete and particles have a partial mixing.  An stronger
separation has been found experimentally \cite{Burtally} in a granular 
mixture of glass and bronze particles vertically shaked. 
These authors found, however, complete segregation.

If we define the transition from BNE and RBNE as the dissipation when 
$Z_A=Z_B$, we find a value for $\alpha_B \simeq 0.71$. In other words, when 
$\sigma_B=1.5$, a dissipation $\alpha_B< 0.71$ makes the $B$ particles to 
sink. As mentioned before, the parameter $\delta$, defined in Eq. 
(\ref{delta}), takes its minimum value: $\delta= 0.053$.

The coexistence of BNE and RBNE could be a transient configuration due, in
part, to the high compacity of the small grains in the middle region that limits the mobility of
the large particles. They could not migrate between the bottom layers and the
top ones. A dynamic analysis, that is beyond the purpose of this article, could
lead some insight into the coexistence of both effects.

\begin{figure}[htb]
\center\includegraphics[angle=270,clip=true,width=0.55\columnwidth,keepaspectratio]
{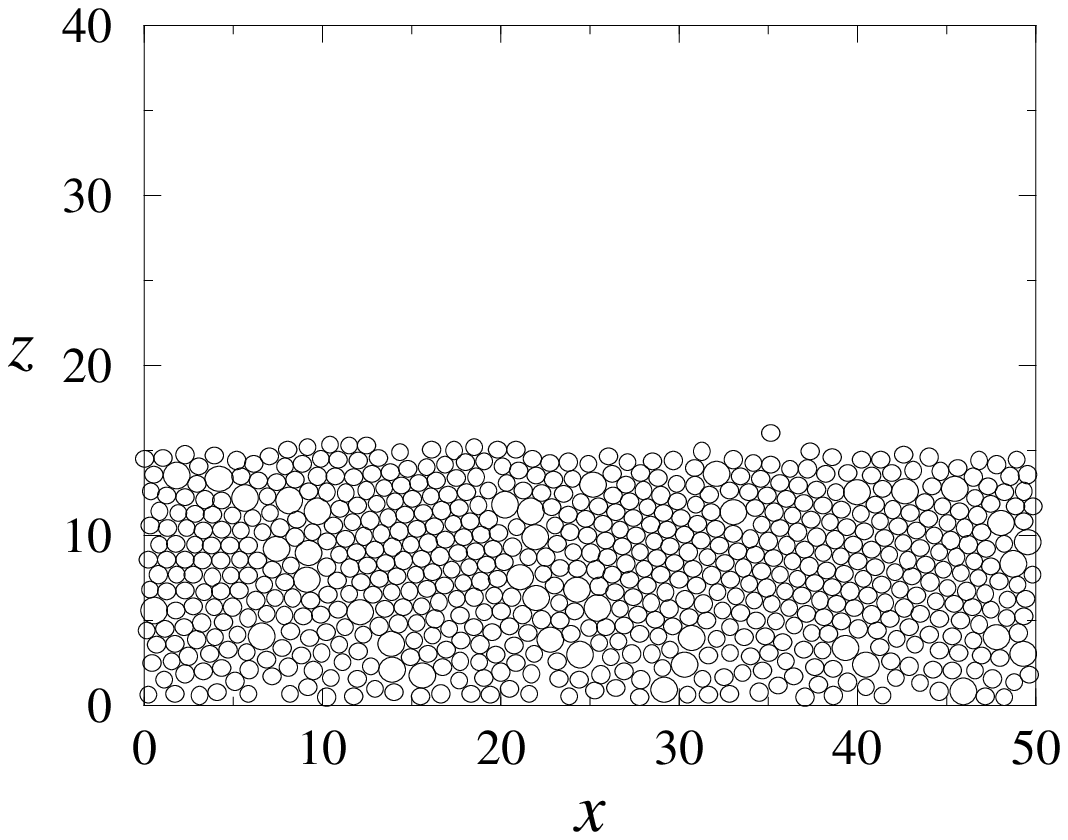}
\caption{Configuration of the grains with $\alpha_A=0.9$, $\alpha_B=0.3$,
$\sigma_B=1.5$, $N_A=640$, and $N_B=40$. The large and more inelastic grains
aggregate in small clusters. }
\label{fig.configRBNE-BNE}
\end{figure}

An inspection of the particle configurations with $\alpha_B=0.3$  where both
RBNE and BNE coexist, shows that $B$ particles develop a tendency to form pairs
and small clusters (see Fig.\ref{fig.configRBNE-BNE}). This phenomenon was
already reported in the case of particles of equal size and it was
quantitatively characterized in terms of the pair distribution functions
\cite{SegregaRestPRE}.
There is not only a macroscopic segregation described in terms of the density
profiles and center of mass positions, but also a microscopic segregation with
a tendency of the more inelastic particles to aggregate in small clusters. In this case, 
the pair distribution function of the $AA$ particles resembles the structure of a hexagonal 
crystaline structure, while the $BB$ pair distribution function has a very high peak 
at $\sigma_B$, followed by a much smaller peak and hardly any structure for distances 
larger than $2\sigma_B$.

\begin{figure}[htb]
\includegraphics[angle=0,clip=true,width=0.95\columnwidth,keepaspectratio]
{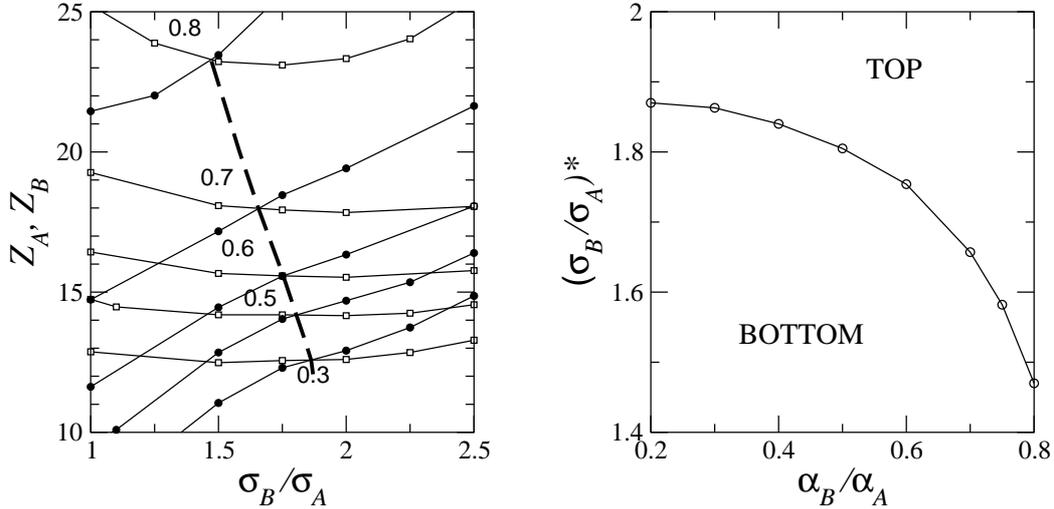}
\caption{Left: Position of the center of mass of $A$ (open squares) and $B$
particles (solid circles) as a function of the relative size\ of $B$
particles
$\sigma_B/\sigma_A$ for different values of $\alpha_B$ (indicated in the plot).
The locus of the mass ratio at which the two centers of mass
coincide is indicated by a thick dashed line.
Right: Critical size ratio $(\sigma_B/\sigma_A)^*$ at which the two centers of
mass coincide as a function of the inelasticity of $B$ particles, $\alpha_B/\alpha_A$.
The regions
in the parameter space where $B$ particles move to the top or to the bottom
are indicated by the respective labels in the plot.
Simulation
parameters are fixed to $\alpha_A=1.0$, $N_A=640$, $N_B=40$, and $m_B=1$.
}
\label{fig.diamratio}
\end{figure}

We have shown that inelasticity induced segregation can compete with
both buoyant forces and the BNE, separately. If a mixture of grains is
considered such that $B$ particles have the same mass as $A$ particles, but have
larger size, both buoyancy forces and the BNE push $B$ particles to the top. The
simulation results presented in Fig. \ref{fig.diamratio} show that for $\sigma_B
\gtrsim 1.9$ buoyancy and BNE cannot be compensated by inelasticity and the
large grains go always to the top. When $\sigma_B \lesssim 1.9$
inelasticity induced segregation can balance the the combined effect of
buoyancy and BNE and higher inelasticities are needed to
compensate their effects for larger $B$ particles.

\begin{figure}[htb]
\center\includegraphics[clip=true,width=0.90\columnwidth,keepaspectratio]
{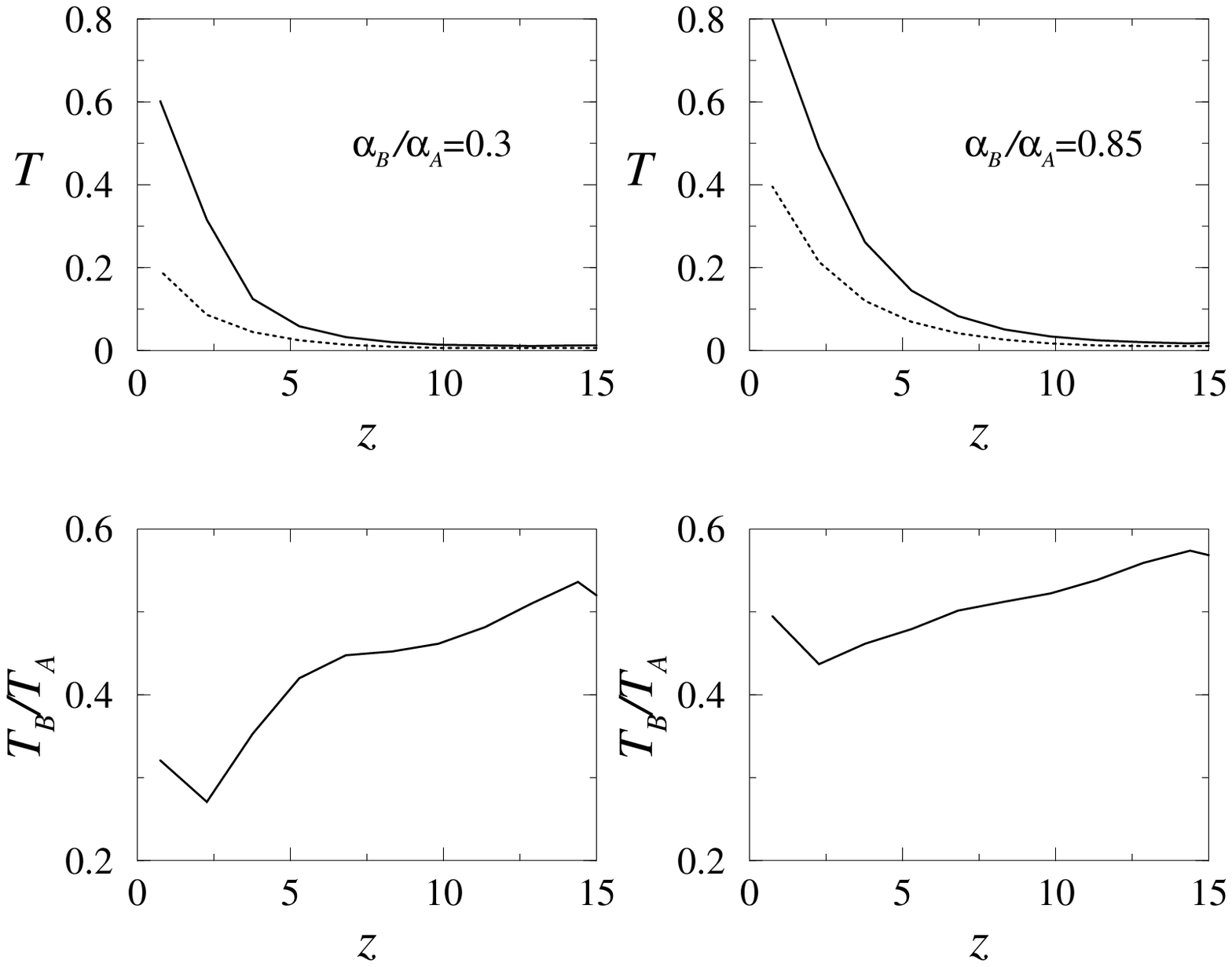}
\caption{Temperature profiles (top) of $A$ particles (solid line) and $B$
particles (dotted line) and temperature ratios $T_B(z)/T_A(z)$ (bottom)  for
$\alpha_B=0.3$ (left) and $\alpha_B=0.85$ (right). Simulations were done with $\alpha_A=0.9$,
$\sigma_B=1.5$, $N_A=640$, $N_B=40$, and $m_B/m_A=2.25$ (neutrally buoyant).}
\label{fig.Temps}
\end{figure}

The temperature profiles present the same qualitative behavior as in the case of
a mixture where grains differ only in their restitution coefficient (see Fig.
\ref{fig.Temps}). That is, an initial abrupt drop, followed by a slight  linear
increase with height. The temperature ratio $T_B(z)/T_A(z)$ shows that $B$
particles are always colder than $A$ particles but it does not
agree with kinetic theory predictions for dilute and moderately dense granular
gases \cite{Brey} as we work in a high density regime, close to close packing in
some of our simulations.

\section{Conclusions} \label{sec.conclusions}
The main conclusion of the present paper is that different restitution
coefficients alone create  segregation in a binary mixture vertically vibrated.
This segregation mechanism can compete or enhance other segregation mechanisms
as buoyancy or the Brazil nut effect. The restitution coefficients
must be considered, in addition to the usual material properties (mass ratio and
diameter ratio), in order to describe accurately the segregation process in a
granular mixture. The effect of the inelasticity is such that the most inelastic
particles sink to the bottom of the container while the less inelastic ones
raise to the top. Segregation is not complete, however, but only partial. The
density profiles of each species show a maximum which are located in
different positions depending on the inelasticities, but they have an
important overlap.

The segregation effects presented here also appears by vibrating a mixture of
elastic and inelastic particles. Again inelastic particles migrate to
the bottom of the container and elastic one prefer the upper part. This
phenomenon allows to make a systematic study of the mechanism reducing the
parameter space.

If, besides the difference of inelasticities, the two species differ in
mass density buoyancy forces enter into play. If the inelastic particles are
less dense, the two mechanism compete and a transition line in the parameter
space is found that distinguish the cases where the inelastic particles have
lower or higher center center of mass compared to the elastic ones. More
inelastic particles need higher buoyancy forces to compensate the
inelasticity induced segregation mechanism. 

When the two species have the same density (that is, they are neutrally
buoyant) but differ in their size, the Brazil nut effect appears: large
particles have the tendency to go to the top. When the large particles are also
more inelastic, the two mechanism compete. The simulations show that there is a
range of parameters where a large number of the large and inelastic grains are
simultaneously at the top and at the bottom: Brazil nut effect and the reverse
Brazil nut effect coexist. The relative proportion of the particles in the two
regions depend on the control parameter.

Concerning the temperatures, most dissipative particles have a lower temperature
than the most elastic ones, both having the characteristic shape of
vibrofluidized system (fast cooling away from the moving boundary followed by a
heating that grows linearly with the distance). Their rate, however, does not
agree with kinetic theory predictions for dilute and moderately dense granular
gases.

We want to thank J.M.R. Parrondo for very useful comments.
R.B. is supported by the Spanish Projects MOSAICO, UCM/PR34/07-15859 and the 
Program Profesores UCM en el Extranjero.
The research is supported by {\em Fondecyt} grants
1061112, 1070958, and 7070301 and {\em Fondap} grant 11980002.

\end{document}